# A Transformer-based Generative Adversarial Network for Brain Tumor Segmentation

Liqun Huang, Long Chen, Baihai Zhang, Senchun Chai

School of Automation, Beijing Institute of Technology, China

**Abstract.** Brain tumor segmentation remains a challenge in medical image segmentation tasks. With the application of transformer in various computer vision tasks, transformer blocks show the capability of learning long-distance dependency in global space, which is complementary with CNNs. In this paper, we proposed a novel transformer-based generative adversarial network to automatically segment brain tumors with multi-modalities MRI. Our architecture consists of a generator and a discriminator, which are trained in min-max game progress. The generator is based on a typical "U-shaped" encoder-decoder architecture, whose bottom layer is composed of transformer blocks with resnet. Besides, the generator is trained with deep supervision technology. The discriminator we designed is a CNN-based network with multi-scale $L_1$ loss, which is proved to be effective for medical semantic image segmentation. To validate the effectiveness of our method, we conducted experiments on BRATS2015 dataset, achieving comparable or better performance than previous state-of-the-art methods.

## 1. Introduction

Semantic medical image segmentation is an indispensable step in computer-aided diagnosis. In the planning of radiation therapy, accurately depicting the area where the tumor is located can maximize the coverage of the target area, and at the same time, it can greatly reduce the toxicity of surrounding high-risk organs. In clinical practice, tumor delineation is usually performed manually or semi-manually, which is time-consuming and labor-intensive. As a result, it is of vital importance to explore automatic volumetric segmentation methods from medical image to accelerate computer-aided diagnosis.

In this paper, we focus on the segmentation of brain tumors with the help of magnetic resonance imaging (MRI) consisting of multi-modality scans. Previous research has shown that, gliomas comprise about 30 percent of brain tumors and central nervous system tumors, and 80 percent of all malignant brain tumors [1]. The automatic segmentation of gliomas remains one of the most challenging medical segmentation problems stemming from some aspects, such as, arbitrary shape and location, poorly contrasted, blurred boundary with surrounding issues.

Since the advent of deep learning, Convolutional Neural Networks (CNN) have achieved great success in various computer vision tasks, ranging from classification, object detection to segmentation. Fully Convolution Networks (FCN [2]) and in particular "U-shaped" encoder-decoder architectures have realized state-of-the-art results in medical semantic segmentation tasks. U-Net [3], which consists of symmetric encoder and decoder, uses the skip connections to merge the extracted features from encoder with decoder at different resolutions, aiming at recovering the lost details during downsampling. Owing to the impressive results in plenty of medical applications, U-Net and its variants have become the mainstream architectures in medical semantic segmentation.

In spite of their prevalence, FCN-based approaches fail to model the long-range dependency, due to its intrinsic limited receptive filed and the locality of convolution operations. Inspired by the great success of transformer-based models in Natural Language Processing (NLP), growing number of researchers propose to apply the self-attention mechanism to medical image segmentation, attempting to overcome the limitations brought by the inductive bias of convolution, so as to extract the long-range dependency and context dependent features. Specially, unlike prior convolution operations, transformers encoder a sequence of patches and leverage the power of self-attention modules to pretrain on large-scale dataset for downstream tasks, like Vision Transformer (ViT [4]) and its variants.

Simultaneously to the Transformers applied in medical image segmentation, Generative Adversarial Networks (GAN), a min-max game, whose core idea comes from the Nash equilibrium of game theory, has revealed excellent performance in medical semantic segmentation. In a typical GAN architecture used for segmentation, GAN consists of two competing networks, a discriminator and a generator. The generator learns the capability of contexture representations, minimizing the distance between prediction and masks, while the discriminator on the contrary maximizes the distance to distinguish the difference of them. The two networks are trained in an alternating fashion to improve the performance of the other. Furthermore, some GAN-based methods like SegAN [5], achieve more effective segmentation performance than FCN-based approaches.

In this paper, we propose a novel transformer-based generative adversarial network for brain tumor segmentation. Inspired by some attempts [6,7] of fusing transformer with 3D CNNs, we design an encoder-decoder generator with deep supervision, where both encoder and decoder are 3D CNNs but the bridge of them is composed of transformer blocks with resnet. Inspired by SegAN [5], we adopt the multi-scale $L_1$ loss to our method with only one generator and one discriminator, measuring the distance of the hierarchical features between generated segmentation and ground truth. Experimental results on BRATS2015 dataset show that our method achieves comparable or better performance.

## 2. Related work

### 2.1 Vision Transformers

The Transformers were first proposed by Vaswani et al. [8] on machine translation tasks and achieved a quantity of state-of-the-art results in NLP tasks. Dosovitskiy et al. [4] then applied Transformers to image classification tasks by directly training a pure Transformer on sequences of image patches as words in NLP, and achieved state-of-the-art benchmarks on ImageNet dataset. In object detection, Carion et al. [9] proposed transformer-based DETR, a transformer encoder-decoder architecture, which demonstrated accuracy and run-time performance on par with the highly-optimized Faster R-CNN on COCO dataset.

Recently, various approaches were proposed to explore the applications of the transformer-based model for semantic segmentation tasks. Chen et al. [10] proposed TransUNet, which added transformer layers to the encoder to achieve competitive performance for 2D multi-organ medical image segmentation. As for 3D medical image segmentation, wang et al. [6] exploited Transformer in 3D CNN for MRI Brain Tumor Segmentation and proposed to use a transformer in the bottleneck of "U-shaped" network on BRATS2019 and BRATS2020 datasets. Similarly,

Hatamizadeh et al. [7] proposed an encoder-decoder network named UNETR, which employed transformer modules as the encoder and CNN modules as the decoder, for the brain tumor and spleen volumetric medical image segmentation.

### 2.2 Generative adversarial networks

The GAN [11] is originally introduced for image generation, making the core idea of competing training with a generator and a discriminator respectively known outside of fixed circle. However, there exists a problem that it is troublesome for the original GAN to remain in a stable state, hence making us cautious to balance the training level of the generator and the discriminator in practice. Arjovsky et al. proposed Wasserstein GAN (WGAN) as a thorough solution of the instability by replacing the Kullback-Leibler (KL) divergence with the Earth Mover (EM) distance.

Various methods were proposed to explore the possibility of GAN in medical image segmentation. Xue et al. [5] used U-Net as the generator and proposed a multi-scale $L_1$ loss to minimize the distance of the feature maps of predictions and masks for the medical image segmentation of brain tumors. Oh et al. [12] took residual blocks into account under the framework of pix2pix [13] and segmented the white matter in FDG-PET images. Ding et al. [14] took an encoder-decoder network as the generator and designed a discriminator based on Condition GAN (CGAN) on BRATS2015 dataset, adopting the image labels as the additional input.

## 3. Methodology

### 3.1 Overall Architecture

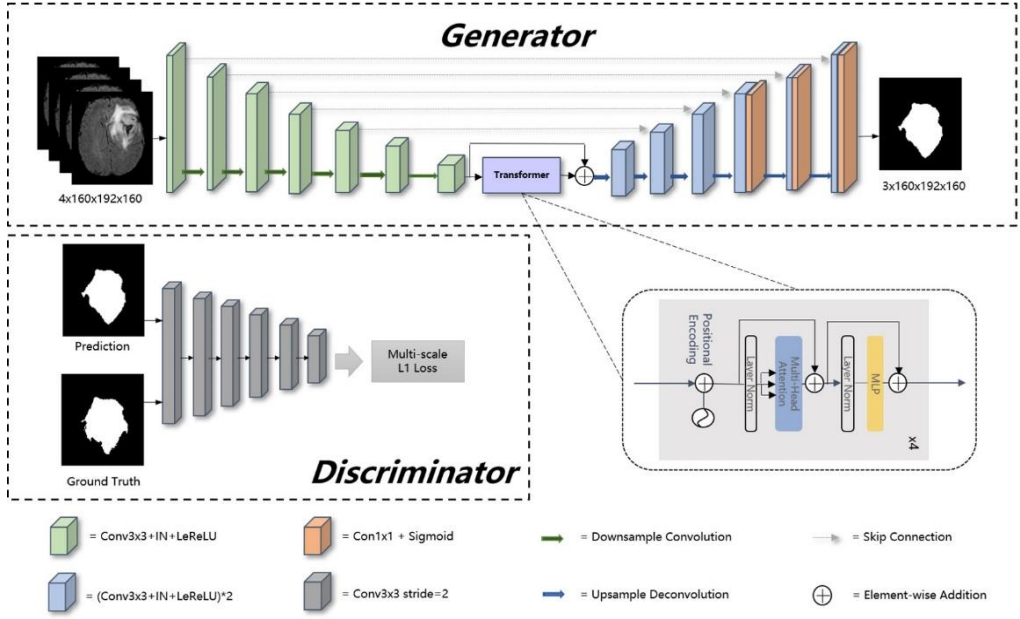

**Figure 1.** Overall architecture of our proposed method.

The overview of our proposed model is presented in Figure 1. Our framework consists of a generator and a discriminator for competing training. The generator G is a transformer-based encoder-decoder architecture. Given a multi modalities (T1, T1c, T2, FLAIR) MRI scan $X \in \mathbb{R}^{C \times H \times W \times D}$ with 3D resolution (H, W, D) and C channels, we utilize 3D CNN-based down-sampling encoder to produce high dimension semantic feature maps, and then these semantic information flow to 3D CNN-based up-sampling decoder through the intermediate Transformer block with resnet [18]. With skip connection, the long-range and short-range spatial relations extracted by encoder from each stage flow to the decoder. For deep supervision [19], the output of decoder consists of three parts: the output of last three convolution layers after sigmoid. Inspired by [5], the discriminator D we used has the similar structure as encoder in G, extracting hierarchical feature maps from ground truth (GT) and prediction separately to compute multi-scale $L_1$ loss.

### 3.2 Generator

Encoder is the contracting path which has five spatial levels. Patches of size 160×192×160 with four channels are randomly cropped from brain tumor images as input, followed by six down-sampling layers with 3D 3×3×3 convolution (stride = 2). Each convolution operation is followed by an Instance Normalization (IN) layer and a LeakyReLU activation layer.

At the bottom of the encoder, we leverage the Transformer block with resnet to model the long-distance dependency in a global space. The feature maps produced by the encoder is sequenced first and then create the feature embeddings by simply fusing the learnable position embeddings with sequenced feature map by element-wise addition. After the position embeddings, we introduce L transformer layers to extract the long-range dependency and context dependent features. Each transformer layer consists of a Multi-Head Attention (MHA) block after layer normalization (LN) and a feed forward network (FFN) after layer normalization. In attention block, the input sequence is fed into three convolution layers to produce three metrics: queries $Q$, keys $K$ and values $V$. To combine the advantages of both CNN and Transformer, we simply short cut the input and output of Transformer block. Thus, as in [8, 6], given the input $X$, the output of the Transformer block with Resnet $Y$ can be calculated by:

$$Y = x + y_L \tag{1}$$

$$y_i = FFN(LN(y_i')) + y_i' \tag{2}$$

$$y_i' = MHA(LN(y_{i-1})) + y_{i-1} \tag{3}$$

$$MHA(Q,K,V) = Concat(head_1, \dots, head_h)W^O \tag{4}$$

$$head_i = Attention(Q,K,V) = softmax(QK^T/\sqrt{d_k})V \tag{5}$$

where $y_i$ denotes the output of $i$th ($i \in [1,2,\dots,L]$) Transformer layer, $y_0$ denotes $X$, $W^O$ are projection metrics, $d_k$ denotes the dimension of $K$.

Unlike the encoder, the decoder uses 3D 2×2×2 transpose convolution for up-sampling, followed by skip connection and two 3D 3×3×3 convolution layers. For a better gradient flow and a better supervision performance, a technology called deep supervision is introduced to utilize the last three decoder levels to calculate loss function. Concretely, we downsampled the GT to the same resolution with these outputs, thus making weighted sum of loss functions in different levels.

## 3.3 Discriminator and Loss function

To distinguish the difference between the prediction and GT, the discriminator D extracts features of GT and prediction to calculate $L_1$ norm distance between them. The discriminator is composed of six similar blocks. Each of these blocks consists of a 3×3×3 convolution layer with a stride of 2, a batch normalization layer and a LeakyReLU activation layer. Instead of only using the final output of D, we leverage the $j$th output feature $f_j^i(x)$ extracted by $i$th ($i \in [1,2,\dots,L]$) layers from image $x$ to calculate multi-scale $L_1$ loss $\ell_D$ as follows:

$$\ell_D(x, x') = \frac{1}{L * M} \sum_{i=1}^{L} \sum_{j=1}^{M} \left\| f_j^i(x) - f_j^i(x') \right\|_1 \tag{6}$$

where $M$ denotes the number of extracted features of a layer in D.

Referring to the loss function of GAN [11], our loss function of the whole adversarial process is described as follows:

$$\min_{\theta_G} \max_{\theta_D} \mathcal{L}(\theta_G, \theta_D) = \mathbb{E}_{x \sim P_{data}} \left( \ell_{bce_{dice}}(G(x)) \right) + \mathbb{E}_{x \sim P_{data}} \left( \ell_D(G(x), y) \right) \tag{7}$$

where $\ell_{bce\_dice}$ denotes that the segmentation maps of generator are used to calculate the BCE loss together with the Dice loss, $x$, $y$ denote the input image and ground truth respectively.

## 4. Experiments

### 4.1 Dataset

In the experiments, we evaluated our method using the Brain Tumor Image Segmentation Challenge 2015 (BRATS2015) dataset. In BRATS2015, there are 220 patient cases in high-grade glioma (HGG) and 55 cases in low-grade glioma (LGG) in the training dataset, which contain manual annotation by clinical experts while 110 patient cases in online testing dataset are provided without annotation. All cases are 3D MRI with four modalities: T1, T1c, T2 and FLAIR. Each modality has the origin size 240×240×155 with the same voxel spacing. The ground truth has five classes: background (label 0), necrosis (label 1), edema (label 2), non-enhancing tumor (label 3) and enhancing tumor (label 4). We divided the 275 training cases into a training set and a validation set with the ratio 9:1 both in HGG and LGG. During training and validation, we padded the origin size 240×240×155 to size 240×240×160 with zeros and then randomly cropped into size 160×192×160, which make sure that the most image content is included.

### 4.2 Implementation Details

Experiments were run on NVIDIA A100-PCIE (4x40GB) system for 1000 epochs using the Adam optimizer [15]. The target segmentation maps are reorganized into three tumor subregions: whole tumor (WT), tumor core (TC) and enhancing tumor (ET). The initial learning rate is 0.0001 and batch size is 4. The data augmentation consists of three parts: (1) padding the

data from 240×240×155 to 240×240×160 with zeros; (2) random cropping the data from 240×240×160 to 160×192×160; (3) random flipping the data across there axes by a probability with 0.5. Both the Dice loss in deep supervision and multi-scale $L_1$ loss are employed to train the network in competing progress. In inference, we converted the transformed three subregions (WT, TC, ET) back to the original labels. Specially, we replace the enhancing tumor with necrosis when the possibility of enhancing tumor in segmentation map is less than the threshold which is chosen according to the online testing scores.

## 4.3 Results

To obtain a more robust prediction, we ensemble ten models trained with the whole training dataset to average the segmentation probability maps. We upload the results of our methods on the BRATS2015 dataset and get the testing scores computed via the online evaluation platform, as listed in Table 1.

Table 1. Performance of some methods on BRATS2015 testing dataset.

| Method | Dice | | | Positive Predictive Value | | | Sensitivity | | |
|---|---|---|---|---|---|---|---|---|---|
| | Whole | Core | Enha. | Whole | Core | Enha. | Whole | Core | Enha. |
| UNET [3] | 0.80 | 0.63 | 0.64 | 0.83 | 0.81 | **0.78** | 0.80 | 0.58 | 0.60 |
| ToStaGAN [14] | **0.85** | 0.71 | 0.62 | 0.87 | **0.86** | 0.63 | 0.87 | 0.68 | 0.69 |
| 3D Fusing [16] | 0.84 | **0.73** | 0.62 | 0.89 | 0.76 | 0.63 | 0.82 | **0.76** | 0.67 |
| FSENet [17] | **0.85** | 0.72 | 0.61 | 0.86 | 0.83 | 0.66 | 0.85 | 0.68 | 0.63 |
| SegAN [5] | **0.85** | 0.70 | **0.66** | 0.92 | 0.80 | 0.69 | 0.80 | 0.65 | 0.62 |
| our method | **0.85** | **0.73** | 0.63 | 0.83 | 0.79 | 0.59 | **0.90** | 0.73 | **0.73** |

Figure 2 shows our qualitative segmentation output on BRATS2015 validation set. This figure illustrates different slices of different patient cases in ground truth and predictions separately.

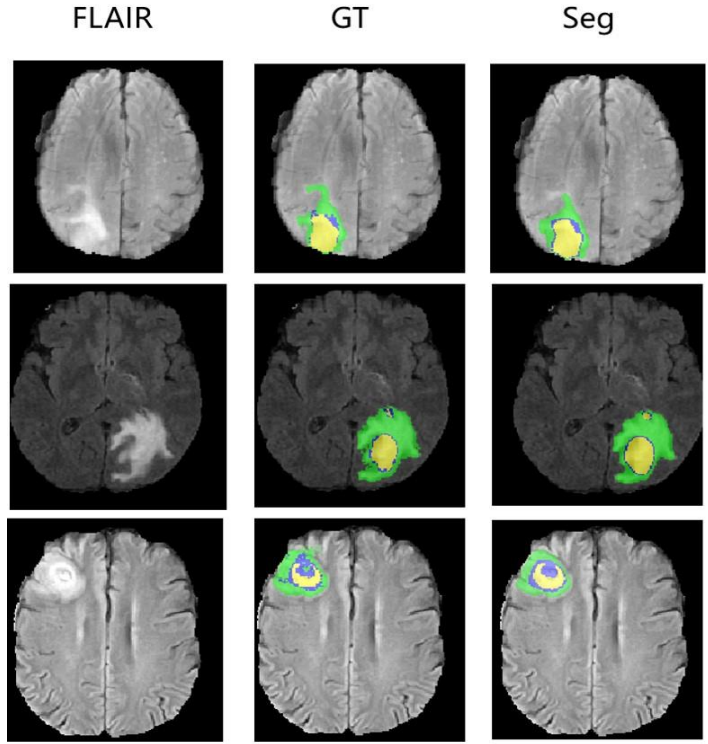

**Figure 2.** Experimental results with corresponding slices on BRATS2015 validation set.

## 5. Conclusion

In this paper, we explored the application of transformer-based generative adversarial network for brain tumor segmentation. Unlike many other encoder-decoder architectures, our generator employs a transformer block with resnet to effectively model the long-distance dependency in a global space, not only inheriting the advantage of CNNs for learning the capability of local contexture representations. Moreover, the application of deep supervision improves the flowability of gradient to some extent. Our discriminator is applied to measuring the norm distance of hierarchical features from predictions and masks. Specially, we calculate multi-scale $L_1$ loss between the generator segmentation maps and ground truth. Experimental results on BRATS2015 dataset show a better performance of our proposed method in comparation of other state-of-the-art methods. In future work, we will apply our method to other medical image segmentation dataset and anticipate that our method performs good generalization.